\documentclass{ws-procs9x6}

\usepackage{psfrag}

\begin{document}

\title{The $\pi NN$ system --- recent progress}

\author{C. Hanhart}

\address{IKP (Theorie), Forschungszentrum J\"ulich,\\
J\"ulich, D--52425, Germany\\
E-mail: c.hanhart@fz-juelich.de\\
www.fz-juelich.de/ikp/theorie/ikpth\_en.shtml}

\begin{abstract}
Recent progress towards an understanding of the  $\pi NN$ system within
chiral perturbation theory is
reported. The focus lies on an effective field theory calculation
and its comparison to phenomenological calculations
for the reaction $NN\to d\pi$. In addition,
the resulting absorptive and dispersive corrections to the $\pi d$
scattering length are discussed briefly.
\end{abstract}


\bodymatter

\section{Introduction}
Pion reactions on few--nucleon systems provide access to
various physics phenomena: deuterons can be used as
effective neutron targets, null--experiments for isospin
violation can be designed, and they are an important test
of our understanding of the nuclear structure. What is
therefore necessary is a controlled theoretical framework
 --- a proper effective
field theory needs to be constructed.

A first step in this direction was taken by Weinberg already
in 1992 \cite{wein}. He suggested that all that needs to be done is to
convolute transition operators, calculated perturbatively in
standard chiral perturbation theory (ChPT), with proper
nuclear wave functions to account for the
 non--perturbative character
of the few--nucleon systems. This procedure
combines the distorted wave born approximation,
used routinely in phenomenological calculations,
with a systematic power counting for the production operators.
Within ChPT this idea was already applied to a large number
of reactions like  $\pi d\to \pi d$ \cite{beane}, $\gamma d\to \pi^0 d$ \cite{kbl,krebs}, $\pi {^3}$He$\to \pi {^3}$He
\cite{baru}, $\pi^- d\to \gamma nn$ \cite{garde}, and $\gamma d\to \pi^+ nn$
\cite{lensky}, where only the most recent references are given.

The central concept to be used in the construction of the
transition operators is that of reducibility, for it allows
one to disentangle effects of the wave functions and those from
the transition operators. As long as the operators are energy
independent, the scheme can be applied straight
forwardly~\cite{wallacephillips},
however, as we will see below, for energy dependent interactions
more care is necessary.

Using standard ChPT especially means to treat the
nucleon as a heavy field. Corrections due to the
finite nucleon mass, $M_N$, appear as contact interactions
on the lagrangian level that are necessarily analytic
in $M_N$.
 However,
some pion--few-nucleon diagrams employ few--body singularities
that lead to contributions non--analytic in $m_\pi/M_N$, with
$m_\pi$ for the pion mass. In Ref. \cite{recoils} it is explained
how to deal with those.

A problem was observed when the original scheme by Weinberg was
applied to the reactions $NN\to NN\pi$ \cite{park,unserd}:
 Potentially higher order corrections turned out to be large and lead
to even larger disagreement between theory and experiment.  For the
reaction $pp\to pp\pi^0$ one loop diagrams that in the Weinberg
counting appear only at NNLO where evaluated \cite{dmit,ando} and they
turned out to give even larger corrections putting into question the
convergence of the whole series.  However, already quite early the
authors of Refs. \cite{bira1,rocha} stressed that an additional new
scale enters, when looking at reactions of the type $NN\to
NN\pi$, that needs to be accounted for in the power counting.
 Since the two nucleons in the initial state need to have
sufficiently high kinetic energy to put the pion in the final state
on--shell, the initial momentum needs to be larger than
$$
p_{thr} = \sqrt{M_Nm_\pi} \ .
$$
The proper way to include this scale was presented
in Ref. \cite{ch3body} and implemented in Ref. \cite{withnorbert} --- for a recent review
see Ref. \cite{report}. As a result, pion $p$-waves are given by tree level diagrams
up to NNLO in the modified power counting
 and the corresponding calculations showed satisfying agreement
with the data \cite{ch3body}. However, for pion $s$--waves loops appear
already at NLO. In the next section we will discuss their effect on the 
reaction $NN\to d\pi$ near threshold. In some detail we will compare
the effective field theory result to that of phenomenological calculations.

Since the Delta--nucleon mass difference, $\Delta$, is numerically of the 
order of $p_{thr}$, also the Delta--isobar should be taken into
account explicitly as a dynamical degree of freedom~\cite{bira1}. We
will use a scheme where 
$$
\Delta \sim p_{thr} \ .
$$

Once the reaction $NN\to d\pi$ is understood within effective field theory
one is in the position to also calculate the so--called dispersive and
absorptive corrections to the $\pi d$ scattering length. This
calculation will be presented in section 3. 

We close with a brief summary and outlook.

\section{$NN\to d\pi$}

\begin{figure}[t!]
\begin{center}
\epsfig{file=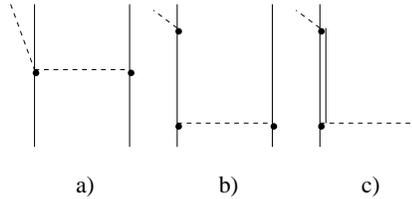,height=2.6cm}
\end{center}
\caption{Tree level diagrams that contribute
to $pp\to d\pi^+$ $s$--waves up to NLO. Solid lines denote 
nucleons, dashed ones pions and the double line 
the propagation of a Delta--isobar.}
\label{tree}
\end{figure}

The tree level amplitudes that contribute to $pp\to d\pi^+$ are shown
in Fig. \ref{tree}.  In Ref. \cite{withnorbert} all NLO contributions
of loops that start to contribute\footnote{In a scheme with two
expansion parameters --- here $m_\pi$ and $p_{thr}$ --- loops no
longer contribute at a single order but in addition to all orders higher than
where they start to contribute.}  to $NN\to NN\pi$ at NLO were
calculated in threshold kinematics --- that is neglecting the
distortions from the $NN$ final-- and initial state interaction and
putting all final states at rest.  At threshold only two amplitudes
contribute, namely the one with the nucleon pair in the final and
initial state in isospin 1 (measured, e.g., in $pp\to pp\pi^0$) and
the one where the total $NN$ isospin is changed from 1 to 0 (measured,
e.g., in $pp\to d\pi^+$)\footnote{The third independent amplitude, where
the $NN$ isospin is changed from 0 to 1 in the production process and
that can be extracted from $pn\to pp\pi^-$, vanishes at threshold as a 
consequence of selection rules.}.  It was found that the sum of all loops that
contain Delta--excitations vanish in both channels.  This was
understood, since the loops were divergent and at NLO no counter term
is allowed by chiral symmetry. On the other hand the nucleonic loops
were individually finite. It was found that the sum of all nucleonic
loops that contribute to $pp\to pp\pi^0$ vanish, whereas the sum of
those that contribute to $pp\to d\pi^+$ gives a finite answer. The
resulting amplitude grows linear with the initial momentum.

\begin{figure}[t!]
\begin{center}
\epsfig{file=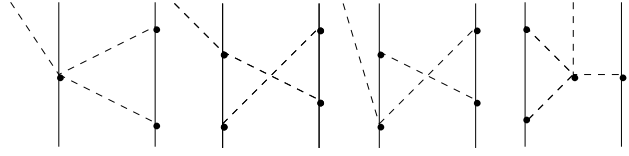,height=2.cm}
\end{center}
\caption{Irreducible pion loops with nucleons
only that start to
contribute to $NN\to NN\pi$ at NLO that were
considered in Ref. \cite{withnorbert}.}
\label{NLOdiags}
\end{figure}

At that time it appeared as a puzzle why the loops vanished for
the reaction $pp\to pp\pi^0$ --- no obvious symmetry reason could
be identified. However, in Ref.~\cite{gep} it was pointed out that
the linear growth of the amplitude for the charged pion production is
the problematic one: when evaluated for finite outgoing $NN$ momenta,
the transition amplitudes turned out to scale as the momentum transfer.
Especially, the amplitudes then grow linearly with the external $NN$ momenta.
As a consequence, once convoluted with the $NN$ wavefunctions, a large
sensitivity to those was found, in conflict with general requirements from
field theory. In light of these insights it was acknowledged that
the loops for $pp\to d\pi^+$ where the ones not understood. The solution
to this puzzle was presented in Ref. \cite{lensky2} and will be reported 
now.

\begin{figure}[t!]
\begin{center}
\psfrag{xx1}{$(m_\pi,\vec 0)$}
\psfrag{xx2}{$(l_0,\vec l)$}
\psfrag{yy1}{$(E+l_0-m_\pi,\vec p+\vec l)$}
\psfrag{yy2}{$(E,\vec p)$}
\psfrag{VV}{\Large $V_{\pi\pi NN}=$}
\epsfig{file=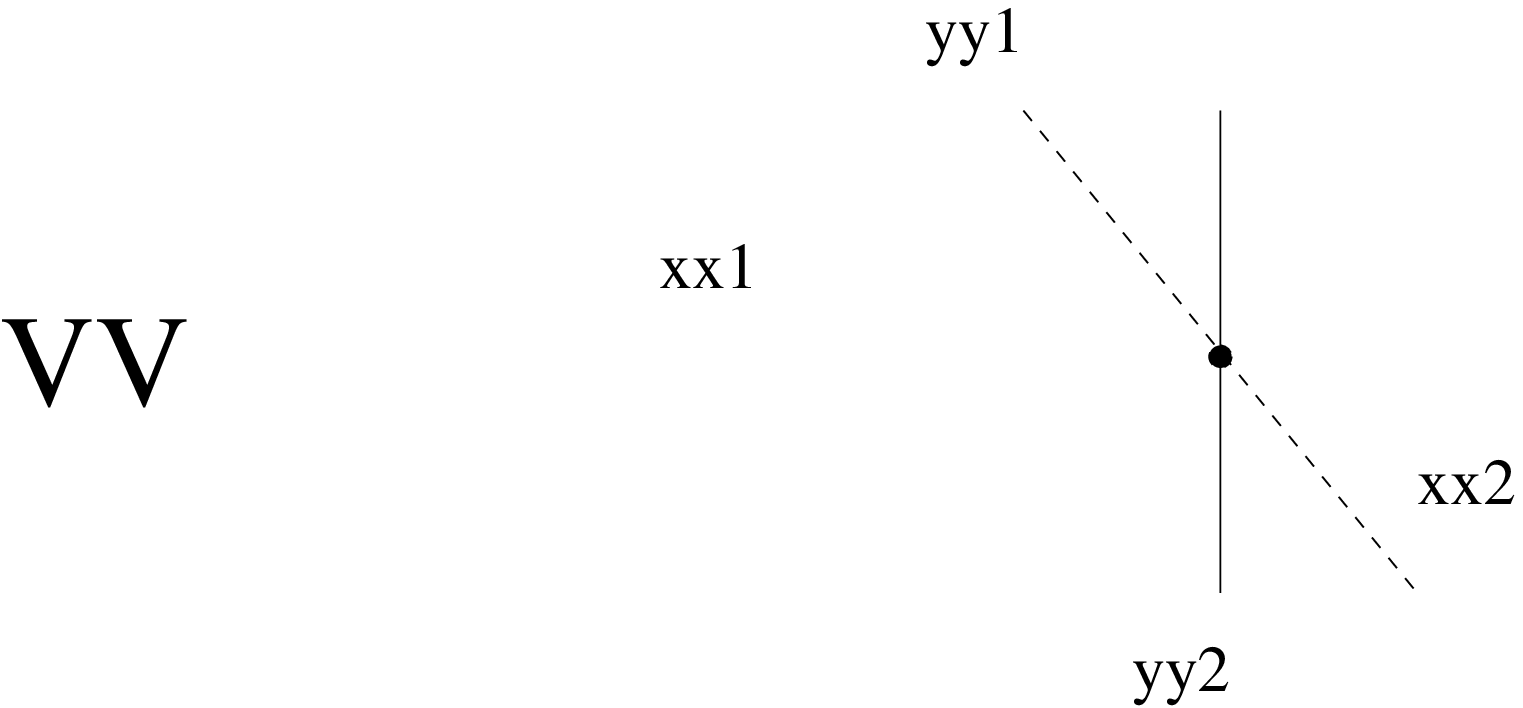, height=2.2cm}
\end{center}
\caption{The $\pi N\to \pi N$ transition
vertex: definition of kinematic variables as used in the text.}
\label{vpipinn}
\end{figure}

The observation central to the analysis is that the leading $\pi N\to
\pi N$ transition vertex, as it appears in Fig. \ref{tree}$a$, is energy
dependent. Using the notation of Fig. \ref{vpipinn} its momentum and
energy dependent part may be written as\footnote{The expressions for
the vertices can be found in Ref.~\cite{ulfs}. Note that the $\pi N\to \pi N$ vertex
from ${\cal L}_{\pi N}^{(1)}$ as well as its recoil correction from  ${\cal L}_{\pi N}^{(2)}$
are to be used already at leading  order as a consequence of the modified power
counting.}
\begin{eqnarray}
V_{\pi\pi NN}&=&
l_0{+}m_\pi{-}\frac{\vec l\cdot(2\vec p+\vec l)}{2M_N} \nonumber \\
&=&
\underbrace{{2m_\pi}}_{\mbox{on-shell}}{+}\underbrace{{\left(l_0{-}m_\pi{+}E{-}
\frac{(\vec l+\vec p)^2}{2M_N}\right)}}_{(E'-H_0)=(S')^{-1}}{-}\underbrace{{\left(E{-}\frac{\vec p\,
    ^2}{2M_N}\right)}}_{(E-H_0)=S^{-1}} \ .
\label{pipivert}
\end{eqnarray}
For simplicity we skipped the isospin part of the amplitude. The first term in
the
last line denotes the transition in on--shell kinematics, the second the
inverse of the outgoing nucleon propagator and third the inverse of the incoming
nucleon propagator. 
First
of all we observe that for on--shell incoming and outgoing nucleons, 
the $\pi N\to \pi N$ transition vertex takes its on--shell value $2m_\pi$ --- even if the
incoming pion is off--shell, as it is for diagram $a$ of Fig. \ref{tree}.
This is in contrast to standard phenomenological treatments\cite{kur}, where 
$l_0$ was identified with $m_\pi/2$ --- the energy transfer in on--shell kinematics ---
and the recoil terms were not considered. Note, since $p_{thr}^2/M_N=m_\pi$ the 
recoil terms are to be kept.

\begin{figure}[t!]
\begin{center}
\epsfig{file=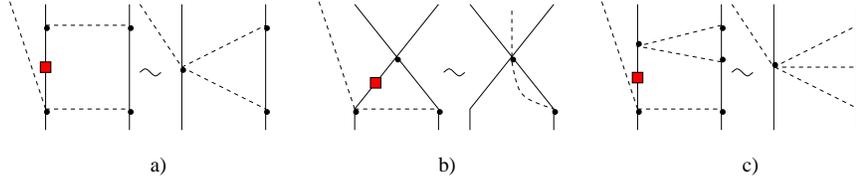, height=2.4cm}
\end{center}
\caption{Induced irreducible topologies, when the
off--shell terms of Eq. (\ref{pipivert}) hit the $NN$ potential
in the final state. The filled box on the nucleon line denotes
the propagator canceled by the off--shell part of the vertex. }
\label{irred}
\end{figure}

The second consequence of Eq. (\ref{pipivert}) is even more interesting: when
the $\pi N\to \pi N$ vertex gets convoluted with $NN$ wave functions, only the
first term leads to a reducible diagram. The second and third term, however,
lead to irreducible contributions, since one of the nucleon propagators gets
canceled.  This is illustrated in Fig. \ref{irred}, where those induced
topologies are shown that appear, when one of the nucleon propagators is
canceled (marked by the filled box) in the convolution of typical diagrams of
the $NN$ potential with the $NN\to NN\pi$ transition operator. Power counting
gives that diagrams $b$ and $c$ appear only at order N$^4$LO and N$^3$LO,
respectively. However, diagram $a$ starts to contribute at NLO and it was
found in Ref. \cite{lensky2} that those induced irreducible contributions
cancel the finite remainder of the NLO loops in the $pp\to d\pi^+$ channel.
Thus, up to NLO only the diagrams of Fig. \ref{tree} contribute to $pp\to
d\pi^+$, with the rule that the $\pi N\to \pi N$ vertex is put on--shell.

\begin{figure}[t!]
\begin{center}
\psfig{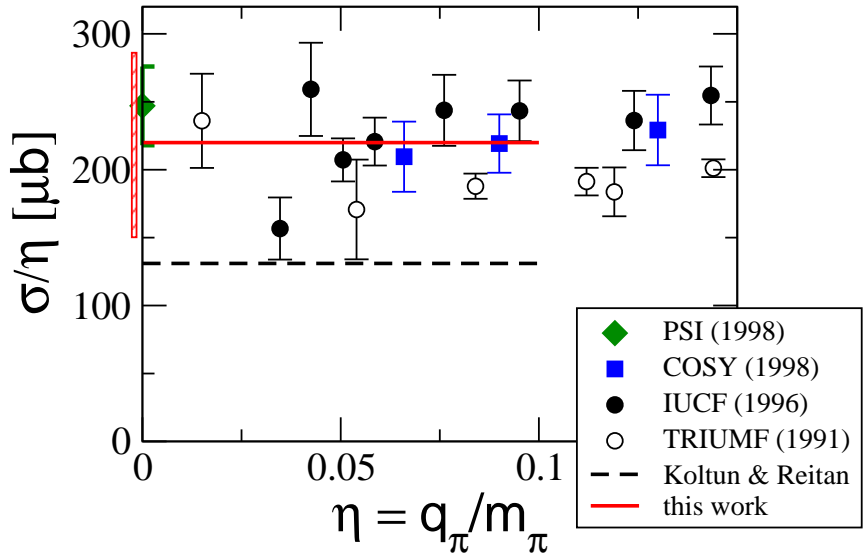}
\end{center}
\caption{Comparison of our results to experimental data for $NN\to d\pi$. 
The dashed line corresponds to the model of Koltun and Reitan~\cite{kur}, 
whereas the solid line is the result of the ChPT calculation of Ref.~\cite{lensky2}.
The estimated theoretical uncertainty (see text) is illustrated by the narrow box.
 The data is from Refs. \cite{dpidata1}
  (open circles), \cite{dpidata2} (filled circles)  and \cite{dpidata3} (filled
  squares). The first data set shows twice the cross section for $pn\to
  d\pi^0$ and the other two the cross section for $pp\to d\pi^+$.}
\label{NNpi_result}
\end{figure}

The result found in Ref. \cite{lensky2} is shown in Fig. \ref{NNpi_result}, where
the total cross section (divided by the energy dependence of phase space) 
is plotted against the normalized pion momentum. The dashed line
is the result of the model by Koltun and Reitan~\cite{kur}, as described
above, whereas the solid line shows the result of the ChPT calculation
of Ref. \cite{lensky2} .

\section{Comparison to phenomenological works}

Phenomenological calculations for the reaction $pp\to d\pi^+$ in near
threshold kinematics are given, e.g., in Ref.~\cite{jouni} and
Ref.~\cite{roleofdel} .  In both works in addition to the diagrams of
Ref.\cite{kur} some Delta--loops as well as short range contributions
are included --- heavy meson exchanges for the former and off--shell
$\pi N$ scattering\footnote{That those are also short range
contributions is discussed in Ref.~\cite{report}.}  for the
latter. Based on this the cross section for $pp\to d\pi^+$ is now even
overestimated near threshold.  How can we interpret this discrepancy
in light of the discussion above?

First of all, the NLO parts of the Delta--loops cancel, as was
shown already in Ref. \cite{withnorbert} . However, in both 
Refs.~\cite{jouni,roleofdel} only one of these diagrams was included
and, especially for Ref. \cite{jouni}, gave a significant contribution.
The only diagram of
those NLO loops shown in Fig.~\ref{NLOdiags} that is effectively
included in Ref.~\cite{roleofdel} is the fourth, since the
pion loop there can be regarded as part of the $\pi N\to \pi N$
transition $T$--matrix. However, as described, the contribution
of this diagram gets canceled by the others shown in Fig.~\ref{NLOdiags}
and the induced irreducible pieces described above. Therefore, the
physics that enhances the cross section compared to the work of Ref.~\cite{kur}
in Refs.~\cite{jouni,roleofdel} is completely different to that
of Ref.~\cite{lensky2} --- the phenomenological calculations
miss the essential contribution and are in conflict with both
field theoretic consistency and chiral symmetry.

What are the observable consequences of the difference
between the ChPT calculation and the phenomenological ones?
As explained, in the former the
near threshold cross section for $pp\to d\pi^+$ is basically given
by a long--ranged pion exchange diagram, whereas the 
latter rely on short ranged operators with
respect to the $NN$ system. Obviously those observables are
sensitive to this difference that get prominent contributions
from higher partial waves in the final $NN$ system. We therefore
need to look at the reaction $pp\to pn\pi^+$. Unfortunately, the total
cross section for this
reaction is largely saturated by $NN$ $S$--waves in the final state (see, e.g.,
Fig.~17 in Ref.~\cite{report}). On the other hand, linear combinations of
 double polarization observables allow one to remove the prominent 
components and the subleading amplitudes should be visible. We
therefore expect from the above considerations that the phenomenological
calculations give good results for polarization observables for $pp\to d\pi^+$,
whereas there should be deviations for some of those for $pp\to pn\pi^+$.
Predictions for these observables were presented in Ref.~\cite{polobs} and
indeed the $\pi^+$ observables with the deuteron in the final state
are described well whereas there are discrepancies for the $pn$ final state
(see Fig.~24 of Ref.~\cite{report}).

It remains to be seen how well the same data can be described in the
effective field theory framework. Up to NNLO the number of counter terms
is quite low: there are two counter terms for pion $s$--waves, that
can be arranged to contribute to $pp\to pp\pi^0$ and $pp\to d\pi^+$
individually, and then there is one counter term for pion $p$--waves,
that contributes only to a small amplitude in charged pion
production~\cite{ch3body}. On the other hand there is a huge
amount of even  double polarized data available~\cite{iucf1,iucf2,iucf3} --- and
there is more to come especially for $pn\to pp\pi^-$~\cite{ankeprop}.

\section{Corrections to $a_{\pi d}$}

The $\pi d$ scattering length is known to a high accuracy from 
measurements on pionic deuterium
 \cite{PSI1}
\begin{equation}
a_{\pi d}^{\mbox{exp}} =\left (-26.1\pm 0.5\mbox \, + \, i (6.3\pm 0.7)\right )\times
\,10^{-3} \ m_\pi^{-1} \ , 
\label{exp}
\end{equation} 
where $m_\pi$ denotes the mass of the charged
pion. In the near future a new measurement with a projected total uncertainty of 0.5\% for
the real part and 4\% for the imaginary part of the scattering length will be
performed at PSI \cite{detlev}.
What is striking with this result is the quite large imaginary part
that may be written as
\begin{equation}
4\pi \mbox{Im}(a_{\pi d})=\lim_{q\to 0}q\left\{
\sigma(\pi d\to NN)+\sigma(\pi d\to \gamma NN)\right\} \ ,
\label{opttheo}
\end{equation}
where $q$ denotes the relative momentum of the initial $\pi d$ pair.
The ratio $R=\lim_{q\to 0}\left(\sigma(\pi d\to NN)/\sigma(\pi d\to
\gamma NN)\right)$ was measured to be $2.83\pm 0.04$
\cite{highland}. At low energies diagrams that lead to a sizable
imaginary part of some amplitude are expected to also contribute
significantly to its real part.
 Those contributions are called dispersive corrections.  As a
first estimate Br\"uckner speculated that the real and imaginary part
of these contributions should be of the same order of magnitude
\cite{brueck}. This expectation was confirmed within Faddeev
calculations in Refs. \cite{at}.  Given the high accuracy of the
measurement and the size of the imaginary part of the scattering
length, another critical look at this result is called for as already
stressed in Refs. \cite{tle,bk} .  A consistent calculation is only
possible within a well defined effective field theory --- the first
calculation of this kind was presented in 
Ref.~\cite{apiddisp} and is briefly sketched here.

\begin{figure}[t!]
\begin{center}
\psfig{file=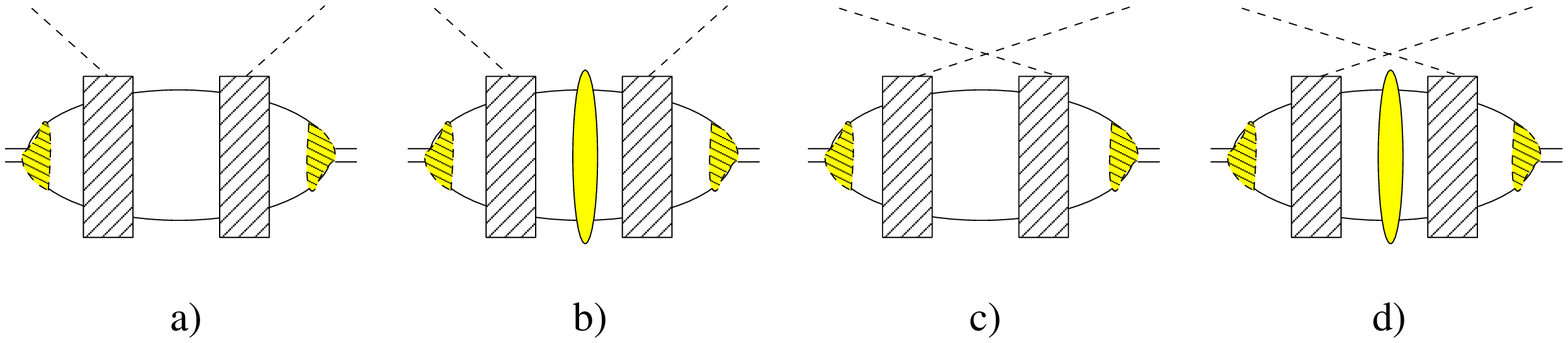,width=4.in}
\end{center}
\caption{Dispersive corrections to the $\pi d$ scattering length.}
\label{disp}
\end{figure}

To identify the diagrams that are to contribute we first need to specify
what we mean by a dispersive correction.
We define dispersive corrections as contributions from diagrams with
an intermediate state that contains only nucleons, photons and at most
real pions. Therefore, all the diagrams shown in Fig.~\ref{disp}
are included in our work. On the other hand, all diagrams
that, e.g., have Delta excitations in the intermediate state
do not qualify as dispersive corrections, although they
might give significant contributions~\cite{doering}.

The hatched blocks in the diagrams of Fig.~\ref{disp} refer
to the relevant transition operators for the reaction $NN\to NN\pi$
depicted in Fig.~\ref{tree}. Also in the kinematics of relevance
here the $\pi N\to \pi N$ transitions are to be taken with
their on--shell value $2m_\pi$. Using the CD--Bonn potential~\cite{cdbonn}
for the $NN$ distortions we found for the dispersive
correction from the purely hadronic transition
\begin{equation}
a_{\pi d}^{disp}=(-6.3+2+3.1-0.4)\times 10^{-3} \, m_\pi^{-1}
=-1.6\times 10^{-3} \, m_\pi^{-1} \ ,
\end{equation}
where the numbers in the first bracket are the individual results
for the diagrams shown in Fig.~\ref{disp}, in order. There are
two points important to stress, first of all the inclusion 
of the intermediate $NN$ interaction is necessary (and required
based on power--counting) and the crossed diagrams (diagram $c$ and $d$) give a 
numerically significant contribution. The latter
finding might come as a surprise on the first glance, however, please
recall that in the chiral limit all four diagrams of Fig.~\ref{disp}
are kinematically identical and chiral perturbation theory is
a systematic expansion around exactly this point. Thus, as a result
we find that the dispersive corrections to the $\pi d$ scattering
length are of the order of 6 \% of the real part of the scattering 
length. Note that the same calculation gave very nice agreement
for the corresponding imaginary part~\cite{apiddisp}.

In Ref.~\cite{apiddisp} also the electro--magnetic contribution
to the dispersive correction was calculated. It turned out that
the contribution to the real part was tiny --- $-0.1\times 10^{-3} \, m_\pi^{-1}$ ---
while the sizable experimental value for the imaginary part (c.f.
Eqs. (\ref{exp}) and \ref{opttheo}) was described well.

To get a reliable estimate of the uncertainty of the calculation just
presented a NNLO calculation is necessary. At that order two counter
terms appear for pions at rest that can be fixed from $NN\to NN\pi$,
as indicated above. For now we need to do a conservative estimate for
the uncertainty by using the uncertainty of order $2\, m_\pi/M_N$ one
has for, e.g., the sum of all direct diagrams to derive a $\Delta
a_{\pi d}^{disp}$ of around $1.4\times 10^{-3} \, m_\pi^{-1}$, which
corresponds to about 6\% of $\mbox{Re}\left(a_{\pi
d}^{\mbox{exp}}\right)$. However, given that the operators that
contribute to both direct and crossed diagrams are almost the same
 and that part of the mentioned cancellations is a
direct consequence of kinematics, this number for $\Delta a_{\pi
d}^{disp}$ is probably too large.

In Ref.~\cite{apiddisp} a detailed comparison to previous
works is given. Differences in the values found for the
dispersive corrections were traced to the incomplete sets
of diagrams included. 

\section{Summary and Outlook}

The process $NN\to NN\pi$ is a puzzle already since more
than a decade. Given the progress presented above we have
now reason to believe that this puzzle will be solved soon.
This mentioned results could only be found, because a consistent
effective field theory was used. For example, the potential problem
with the transition operators of Ref.~\cite{withnorbert},
pointed at in Ref.~\cite{gep}, would
always be hidden in phenomenological calculations, since the
form factors routinely used there always lead to finite, well behaved
 amplitudes.
The very large number of observables available for
the reactions $NN\to NN\pi$ will provide a non--trivial test
to the approach described.

Once the scheme is established, the same field theory can
be used to analyze the isospin violating observables measured
in $pn\to d\pi^0$~\cite{opper} and $dd\to \alpha\pi^0$~\cite{ed}.
First steps in this direction were already done in Ref.~\cite{knm}
for the former and in Refs.~\cite{csb1,csb2} for the latter.

Based on the calculation for $pp\to d\pi^+$ we also performed a
calculation for the dispersive and absorptive corrections to the
$\pi d$ scattering length that were calculated for the first time
within ChPT. The final answer turned out to be relatively small as
a consequence of cancellations amongst various terms. This work
is an important step forward towards a high accuracy calculation
for the $\pi d$ scattering length that will eventually allow for
a reliable extraction of the isoscalar scattering length. However,
before this can be done, isospin violating corrections~\cite{mrr} as well
as the contributions from the Delta--isobar need to be evaluated.

\section*{Acknowledgments}
I thank the organizers for a very well organized
and educating conference and V. Lensky, V. Baru,
J.~Haidenbauer, A. Kudryavtsev, and U.-G. Mei\ss ner for a very
fruitful collaboration that lead to the results presented.  This
research is part of the EU Integrated Infrastructure Initiative Hadron
Physics Project under contract number RII3-CT-2004-506078, and was
supported also by the DFG-RFBR grant no. 05-02-04012 (436 RUS
113/820/0-1(R)) and the DFG SFB/TR 16 "Subnuclear Structure of
Matter".  A.~K. and V.~B. acknowledge the support of the Federal
Program of the Russian Ministry of Industry, Science, and Technology
No 02.434.11.7091.

 \end{document}